\documentclass[10pt,twocolumn]{article}

\usepackage{graphicx}
\usepackage{amsfonts}
\usepackage{amsmath}
\usepackage{amssymb}
\usepackage{mathrsfs} 
\usepackage{lipsum}
\usepackage{color}
\usepackage{siunitx}
\usepackage{authblk}
\usepackage[margin=2cm]{geometry}
\usepackage{mathtools, cuted}

\usepackage{bm}
\usepackage{amsmath}

\title{Protocol to perform pressurized blister tests on thin elastic films}
\author[1,2]{Fran\c{c}ois Boulogne}
\author[1]{Sepideh Khodaparast}
\author[2]{Christophe Poulard}
\author[1]{Howard A. Stone}

\affil[1]{Department of Mechanical and Aerospace Engineering, Princeton University, Princeton, NJ 08544, USA}
\affil[2]{Laboratoire de Physique des Solides, CNRS, Univ. Paris-Sud, Universit\'e Paris-Saclay, Orsay 91405, France}

\date{\today}
\begin{document}

\twocolumn[
    \begin{@twocolumnfalse}
        \maketitle
        \begin{abstract}
This work aims to identify common challenges in the preparation of the blister test devices designed for measurement of energy release rate for brittle thin films and to propose easy-to-implement solutions accordingly.
To this end, we provide a step-by-step guide for fabricating a blister test device comprised of thin polystyrene films adhered to glass substrates.
Thin films are first transferred from donor substrates to an air-water interface, which is then used as a platform to locate them on a receiver substrate.
We embed a microchannel at the back of the device to evacuate the air trapped in the opening, through which the pressure is applied.
We quantify the height and the radius of the blister to estimate the adhesion energy using the available expressions correlating the normal force and the moment with the shape of the blister.
The present blister test provided adhesion energy per unit area of $G = 18 \pm 2$ $ \mbox{mJ}/{\mbox m}^2$ for polystyrene on glass, which is in good agreement with the measurement of $G = 14 \pm 2$ $ \mbox{mJ}/{\mbox m}^2$ found in our independent cleavage test.
        \end{abstract}
    \end{@twocolumnfalse}
]

%
%
\section{Introduction}\label{sec:intro}

Characterization of adhesion energies between materials is of broad interest for scientific and engineering purposes.
In practice, measuring the work of adhesion remains a difficult task both for the complex mechanical problems involved and for the technical barriers.
Several measurement techniques have been developed in the past century.
The most intuitive test is probably the peeling test \cite{Obreimoff1930,Kendall1975} that aims to propagate an interfacial crack between two materials.
However, certain materials are difficult to manipulate as required in the peeling test, especially when the film is particularly thin and brittle.
In addition, the Johnson-Kendall-Roberts contact adhesion test relies on the deformation of a small sphere in contact with a surface of interest \cite{Johnson1971}, which assumes thick materials to avoid finite-thickness effects \cite{Shull1997,Shull2002,Barthel2008}.

Dannenberg developed the blister test in 1961 to measure the adhesion of paints on surfaces \cite{Dannenberg1961} which was further developed in particular by Jensen \cite{Jensen1991,Jensen1998}.
The so-called blister test consists of debonding a thin film by imposing a pressure via injecting a fluid between the film and the substrate to form a blister.
Since then the blister test has been employed in different systems to quantify the adhesion energy between films of different properties and thicknesses and a large variety of substrates \cite{Hinkley:1983,Briscoe1991,Sizemore1995,Dupeux:1998,Cao:2014} including graphene membranes \cite{Koenig:2011,Huang:2011,Boddeti:2013,Metten:2014}.
Originally, the pressure and the volume of the injected liquid were measured to estimate the work required to detach the film \cite{Dannenberg1961}.
However, recent studies have often employed more advanced optical techniques such as interferometry to quantify the shape of the blister \cite{Cao:2014,Yahiaoui:2001}.

Multiple studies have investigated the theoretical and empirical models to correlate the adhesion energy with the shape of the blister for the potential different deformation regimes in a blister test, including the bending plate, the stretching membrane and the transitional regime in between \cite{Guo:2005,Komaragiri:2005,Wang:2016}.
Recently, Sofla \textit{et al.} (2010) proposed a mechanical model, which was solved numerically, to relate the energy release rate to the morphology of the deformed film in a blister test for a wide range of physical parameters, which covers both the membrane and the plate regimes \cite{Sofla2010}.
These authors quantified the adhesion energy of millimeter thickness polydimethylsiloxane PDMS films on glass.
However, for thinner and more brittle materials, this measurement can be significantly more difficult to carry out.
In particular, in the present study we show the preparation and the measurement protocols for polystyrene films with higher elastic modulus ($E = 3.4$~GPa) compared to less stiff materials such as PDMS ($E \approx 1.2$~MPa), which imposes additional challenges in the preparation and transfer of the thin film, and visualization of the deflection of the blister.

\begin{figure*}[htp]
\center
\resizebox{0.8\textwidth}{!}{
  \includegraphics{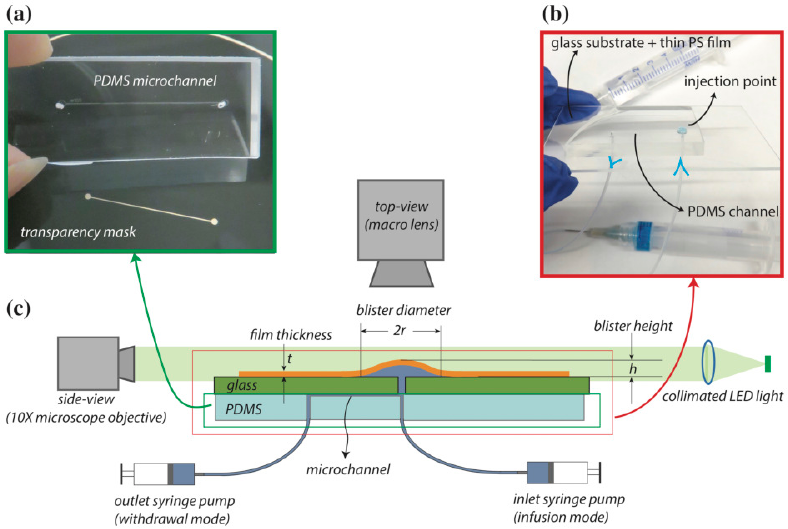}}
\caption{Schematic of the setup for the blister test. (a) A microchannel with a cross-sectional area of $250 \times 250$ $\mu {\rm m}^{2}$ is fabricated in PDMS using a transparency mask and performing soft lithography technique. (b) The inlet of the PDMS microchannel is aligned with the hole drilled in the receiver substrate before the two pieces are bonded together. Finally, the thin PS of thickness $t$ is transferred to the free side of the glass slide to create the blister test device. (c) Dyed water is injected through the inlet underneath the thin film, whilst the trapped air in the hole is evacuated through the outlet of the microchannel. The side-view and top-view images of the blister are captured to determine the maximum deflection $h$ and the diameter $2r$ of the blister, respectively.}
\label{fig:blister_test}
\end{figure*}

Although preparation of the blister device and performing the tests are often described to be rather simple tasks, multiple practical challenges are faced during the process, which are rarely discussed in earlier works.
In fact, most of the previous studies are dedicated more to the estimation of the work of adhesion using the blister test than to how to perform the test.
Thus, in this paper, we aim to focus on how to prepare sample devices and perform the blister test itself, especially for thin elastic films.
We illustrate our method on polystyrene films with thicknesses that range between $590$ to $1200$~nm.
This protocol includes the preparation of a specific device to pressurize a film and the coating of this device with the film of interest.
To validate our experimental approach and the numerical model used to derive the energy release rate, we compared our results to those obtained with a cleavage test on a polystyrene film of thickness about $100$~$\mu$m.

\section{Preparation of the blister test device}\label{sec:device}

In this section, we describe the preparation of a device to perform the blister test depicted in Fig.~\ref{fig:blister_test}, which aims to measure the energy release rate between a thin film and a substrate.
The device must consist of a flat substrate made of the material of interest and pierced in its center for the inlet.
The surface must be covered by the thin elastic film, which is then pressurized by injecting a liquid through the inlet to perform the blister test measurement.
The shape of the blister formed by the thin film can be related to the energy release rate.

To compensate for the high elastic modulus of materials such as polystyrene, adhesion energy measurements of films must be performed on small thicknesses to avoid crack propagation over large distances.
Therefore, the inherent difficulty of the blister test is to make a thin film on a pierced surface, which prevents the use of spin-coating techniques.

The principle of our protocol is to spin-coat a solution of polystyrene on a donor substrate, and the resulting thin film is then floated on a water bath.
We prepare a receiver substrate of interest, pierced in its center, which is connected to an inlet for injecting the liquid.
Then, the film is transferred to the latter surface with a method inspired by the Langmuir-Blodgett (LB) technique \cite{Blodgett1937}.

A second difficulty is to withdraw the air trapped in the injection hole.
Having a blister filled only with the liquid phase is of critical importance, especially if the shape of the blister is estimated according to the volume of the liquid \cite{Hohlfelder1996}.
Here, we propose to add a microchannel to the back of the blister test device for this purpose, as shown in Fig. \ref{fig:blister_test}.

\subsection{Donor substrates}\label{subsec:donor}
Microscope glass slides (Dow Corning, 25 mm by 75 mm) are used as donor substrates in all experiments.
To clean the surface of the glass slides, we plunge them in a bath of acetone for 30 minutes and then they are thoroughly rinsed with deionized (DI) water, and a solution of ethanol and acetone.
Glass slides are dried with clear air and heated at $100^{\circ}$C for 30 minutes prior to the experiments.

\subsection{Preparation of PS films on donor substrates}\label{subsec:psfilm}
\begin{figure}
 \center
 \resizebox{0.35\textwidth}{!}{
 \includegraphics{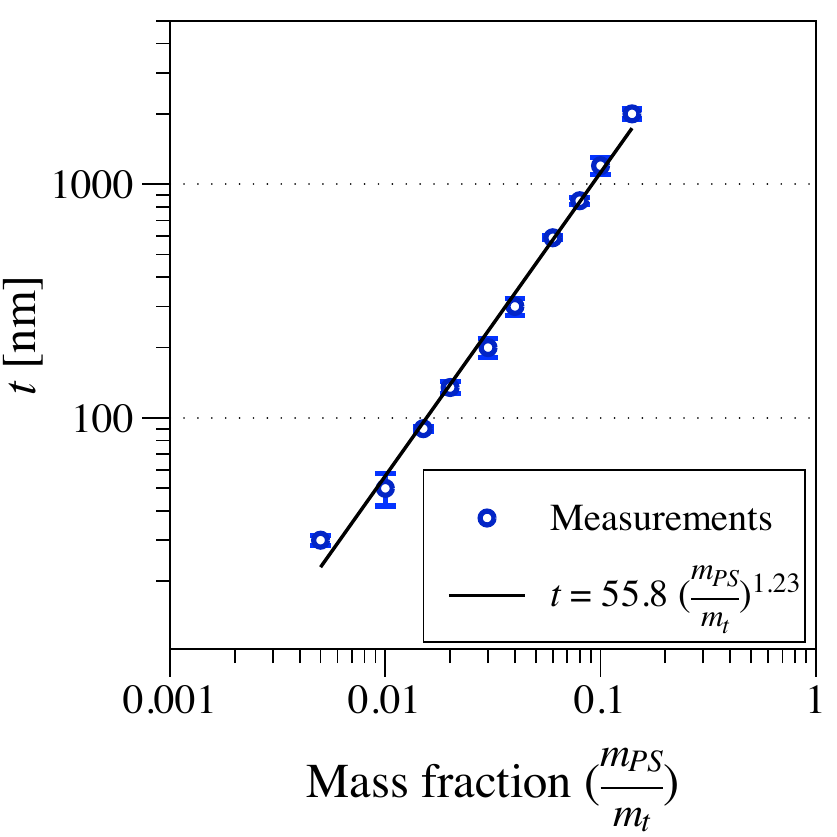}}
 \caption{Thickness $t$ of the PS films vs. the mass fraction of PS in toluene used in the spin-coating.
Error bars represent the standard deviations of the measurements.}
 \label{fig:PS-film}
 \end{figure}

Polystyrene PS films are produced from a solution of PS (Sigma-Aldrich, $M_w \simeq 280$ kg/mol) in toluene.
To achieve thin films of uniform thickness, the PS solution is spin-coated on solid glass substrates at 2000 rpm for 30 seconds.
Solutions of different mass fractions of PS in toluene ($0.005 < \frac{m_{PS}}{m_t} < 0.14$) are used to achieve film thicknesses ranging from $t = 50$ nm to $t = 2$ $\mu$m.
A frame is then cut out with a sharp blade around the spin-coated films before they were annealed at $130^{\circ}$C under vacuum for two hours to release any pre-stress in the films.
After annealing, the thickness of the polystyrene films are measured by a Leica DCM 3D optical profilometer.
To ensure the spatial uniformity of the film thickness for the polystyrene films, we measured the film thickness in four different locations of the film using a Woollam M2000 Spectroscopic Ellipsometer to confirm the values obtained with the optical profilometer.
The average film thicknesses $t$ achieved at different polystyrene concentrations are presented in Fig. \ref{fig:PS-film}.
The resulting films have an elastic modulus $E = 3.4 $~GPa and Poisson's ratio $\nu = 0.33$ \cite{Brandrup1989}.
The measurement of the elastic modulus can be performed with different techniques such as indentation \cite{Oliver2004} or buckling instability \cite{Stafford2004}.

\subsection{Receiver substrates}
Microscope glass slides (Dow Corning, 25 mm by 75 mm) are used as receiver substrates.
A hole of approximately 1 mm diameter is drilled in the middle of the glass slide with a 1 mm diameter diamond drill bit mounted on a Dremel tool.
The receiver glass slides are cleaned after the drilling process following the procedure described above for the donor substrates.

A microchannel of length $L = 20$ mm and with a cross-section of $250 \times 250$ $\mu {\rm m}^{2}$ is fabricated in PDMS (Dow Corning, Sylgard 184 at a 1:10 wt ratio of crosslinking agent to prepolymer) using conventional soft lithography techniques \cite{Xia:1998}.
The lithography mask used in our experiments is provided in Supplementary Materials.
As shown in Fig. \ref{fig:blister_test}a, two vertical openings of 1 mm diameter are created in the PDMS slab at the inlet and the outlet of the channel with a biopsy punch (Miltex, 98PUN6-1).
The PDMS microchannel and the receiver glass slide are then activated with a plasma gun (Electro-Technic Product, BD20-AC) \cite{Haubert2006}.
Following this step, the inlet of the microchannel is aligned with the hole  drilled in the receiver glass slide, and the two activated pieces are bonded together.
In order to enhance the bonding, the assembly is heated at $90^{\circ}$C for 1 hour (Fig. \ref{fig:blister_test}b).

\subsection{Transferring the PS film}
The thin PS film is transferred from the donor substrate to the receiver substrate.
The first step consists of detaching the PS film from the donor surface.
To this end, the donor substrate is slowly dipped into a water bath until the entire PS film is floating on the free air-water interface.
The water bath is then placed on a motorized translation stage (Thorlabs, NRT200) with the receiver glass slide dipped in it.

\begin{figure}
\centering
\includegraphics{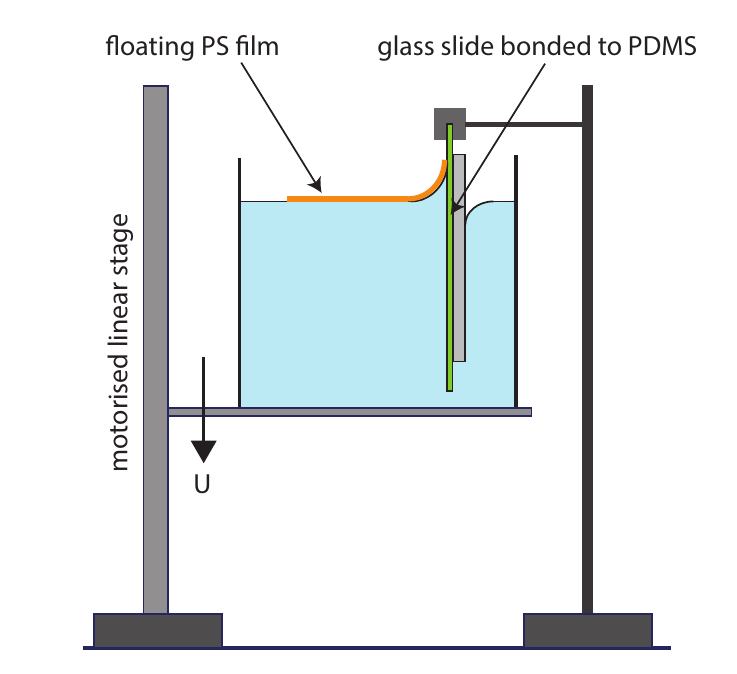}
\caption{Schematic of the experimental setup for the transfer of the PS film (orange) onto the device.}
\label{fig:transfer}
\end{figure}

The second step in the transfer process starts with a PS film floating at the water interface.
The thin floating PS film is brought in contact with the glass side of the receiver substrate and the film slightly bends due to the curvature of the meniscus.
Simultaneously, the water bath is displaced downward at a speed $U = 1$ $\mu$m/s (Fig. \ref{fig:transfer}), to mimic the LB deposition technique \cite{Roberts:1981}.
The air-water interface, which holds the thin PS film moves towards the hydrophilic glass slide and subsequently transfers the thin film onto the substrate \cite{Savelski:1995,Cerro:2003}.

The film transfer is easier for small water contact angles on the receiving glass slide.
Optionally, a stripe of one centimeter at the top of the receiver substrate can be activated by a plasma treatment.
During this treatment, the rest of the glass slide is covered with another glass slide to prevent any modification of the surface that would affect the adhesion energy.

Finally the prepared device is annealed at $130^{\circ}$C under vacuum overnight to release the eventual pre-stress in the films and remove water trapped underneath the film.
The pre-stress could be caused by the mechanical vibrations of the translation stage occurring during the film transfer.
At the end of this process, we obtain a device coated with a PS film ready for the blister test (Fig. \ref{fig:blister_test}b).

\section{Measurement of the adhesion energy in a blister test}\label{sec:blister_test}

Devices are mounted with the PS film facing up on a 3-axis translational stage (Thorlabs, PT3) and a rotational stage (Thorlabs, PR01) equipped with micrometers to ensure the alignment of the device with a horizontal microscope.
The inlet and outlet of the microchannel on the PDMS side of the device are connected to two syringes, which are mounted on two syringe pumps (Harvard Apparatus, PhD Ultra) functioning in infusion and withdrawal modes, respectively.
Fig. \ref{fig:blister_test}c shows the schematic of the experimental setup used in the present blister test measurements.
For visualization purposes, the inlet syringe is filled with a solution of a dye (Sigma-Aldrich, methylene blue) dissolved in DI water.
The microchannel is slowly filled with water by pushing the liquid through the inlet while removing the trapped air in the channel through the outlet.
At this point the blister test device is ready for measurements.

A finite volume of liquid is pumped through the hole in the glass substrate until an interfacial crack is initiated between the two materials.
The system is left to achieve equilibrium at this stage, especially by inspecting that the radius of the bulge does not further evolve.
Once a blister is formed underneath the PS film, the radius and the height of the blister are measured by top-view and side-view visualization of the blister, respectively.

\subsection{Visualization of the blister shape}
The side-view visualization is performed using a horizontal home-made microscope.
The microscope consists of a long working distance infinity corrected apochromatic objective (Mitutoyo 10$\times$, focal length $20$ mm) aligned to a tube lens (Mitutoyo, MT-1, $1\times$) with a  $57$ mm long tube (2.5 cm diameter).
The tube lens is mounted on a DSLR camera (Nikon, D7100) with a 161 mm long tube and a C-mount-to-Nikon lens mount adapter (Fotodiox).
This optical setup provides a spatial resolution of 0.5 $\mu$m per pixel.
A green LED light source (Thorlabs, M530L3) is positioned in line with the tube microscope and behind the blister test device.
The LED light is collimated using a biconvex lens (Thorlabs, LB1761) located at its focal distance from the LED light source.

\begin{figure}
\centering
\resizebox{0.45\textwidth}{!}{
\includegraphics{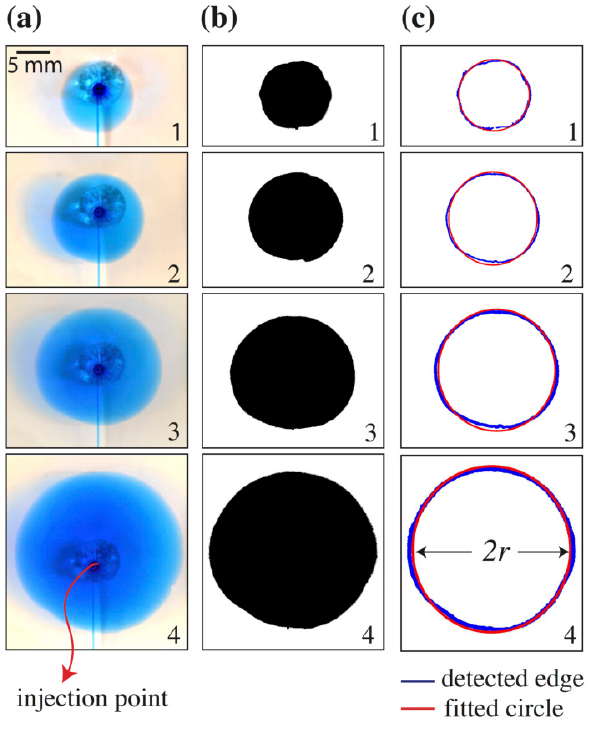}}
\caption{Top-view visualization of the infused liquid between the polystyrene film and the glass substrate for four liquid volumes. (a) Original image, (b) thresholded image and (c) contour detection and fitted circle.}
\label{fig:top_view}
\end{figure}

\begin{figure}
\centering
\includegraphics{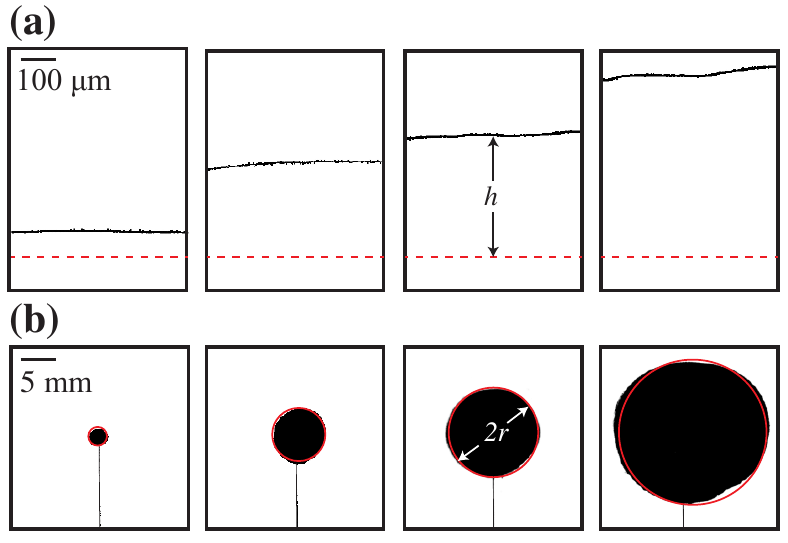}
\caption{Visualization of the deflection of the film. (a) Side-view visualization of the center of the blister. The dashed red line represents the zero level before the injection of liquid under the PS film.
(b) The corresponding top-view images are included to facilitate visualizing the size of the blister.}
\label{fig:side_view}
\end{figure}

The top-view visualization, which provides the diameter $2r$ of the blister, is performed with a macro lens (Nikon, 105 mm) mounted on a DSLR camera (Nikon, D5100).
The spatial resolution in the top-view optical system is 10 $\mu$m per pixel.
Sample snapshots of the top-view and the side-view visualizations are provide in Fig. \ref{fig:top_view} and Fig. \ref{fig:side_view}, respectively.
The diameter of the blister $2r$ is quantified by fitting a circle to the thresholded image of the blister from the top.
A Python code performing this task with Scikit-image \cite{Vanderwalt2014} is provided in the Supplementary Materials.
The height is simply measured in the side-view images by identifying the distance between initial level of the thin film and the highest level of the blister for every injection volume.

\begin{figure}
\centering
\resizebox{0.35\textwidth}{!}{
\includegraphics{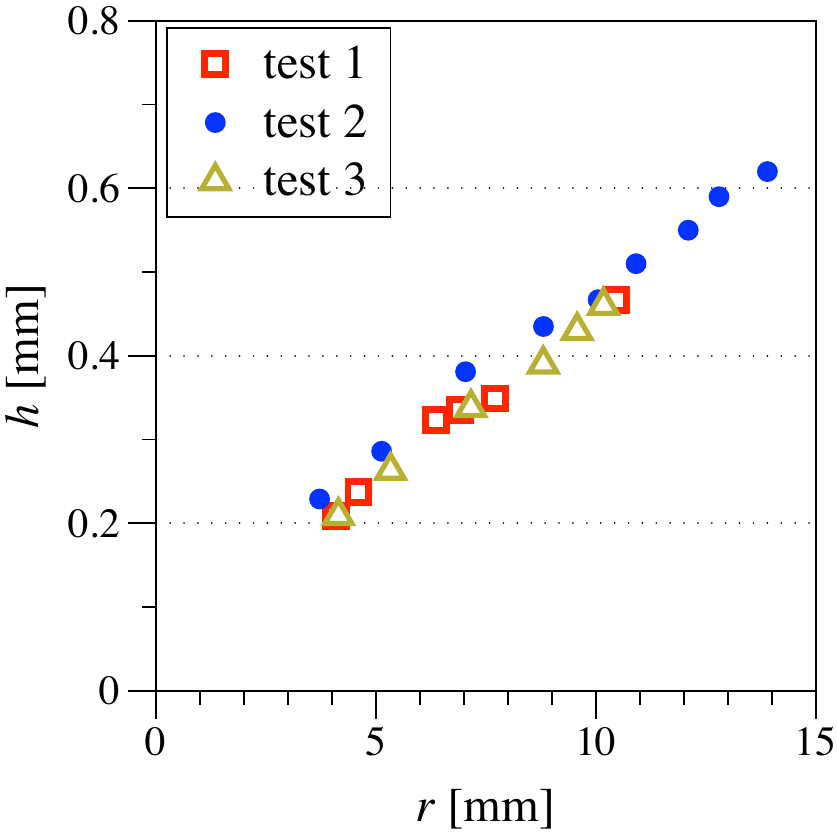}}
\caption{The height versus the radius of the blister as quantified in the experimental measurements in three separate tests for a film thickness $t = 850$ nm.
The maximum relative errors for measurements of the height $h$ and radius $r$ of the blister were $2\%$ and $5\%$, respectively.}
\label{fig:reproducibility}
\end{figure}

The reproducibility of the measurements is controlled by performing measurements on three different devices.
As an example, Fig.~\ref{fig:reproducibility} shows the measurements of the height of the blister versus its radius obtained on three devices for a film thickness $t = 850$ nm.

\subsection{Estimation of the adhesion energy}

Blister tests measurements are performed for three different film thicknesses, namely $t = 590, 850, 1200$ nm.
To derive the energy release rate $G$ as a function of the blister height $h$, we use the expressions provided by Sofla \textit{et al.} \cite{Sofla2010}

\begin{equation}\label{eq:adhesion_buldge}
	G = \frac{\bar E t^5}{2 r^4} \left( 12 \bar M(h)^2 + \bar N(h)^2 \right)
\end{equation}
where $\bar M(h)$ and $ \bar N(h)$ are the dimensionless moment per unit length and the normal force, respectively and $\bar E = E / (1 - \nu^2)$.
The expressions for $\bar M(h)$ and $\bar N(h)$ are given in the Appendix.
The average values of adhesion energy $G$ obtained in the blister tests are presented in Table~\ref{tbl:blister_adhesion}.
The reported results and the corresponding errors indicate the average and standard deviation achieved for individual tests in three separate devices.
The agreement of these values of $G$ for different thicknesses shows the self-consistency of the method.

\begin{table}
  \caption{Measurements of adhesion energy using blister test.
    The uncertainty on the film thickness is calculated from the standard deviation of the thicknesses measured on a batch of films obtained under the same experimental conditions.
    }
  \label{tbl:blister_adhesion}
\begin{tabular}{c c}
\textbf{Film thickness, $t$ [nm]} & \textbf{Adhesion energy, $G$ [mJ/${\mbox m}^2$]} \\
  \hline
    \hline
{\footnotesize 590 $\pm$ 10} & {\footnotesize 18.5 $\pm$ 2.5} \\
    \hline
{\footnotesize 850 $\pm$ 30} & {\footnotesize 17.8 $\pm$ 2} \\
    \hline
{\footnotesize 1200 $\pm$80} & {\footnotesize 18.6 $\pm$ 2} \\
    \hline
\end{tabular}
\end{table}

\section{Comparison with a cleavage test}\label{sec:result}

To validate the analysis made on the blister test, we measured the adhesion energy with a second method, namely a cleavage test, on the same material.
The cleavage test consists of propagating an interfacial crack between the film and the substrate with a wedge.
As the wedge must be pushed between the film and the substrate, this test prevents us to measure the adhesion energy on thicknesses smaller than few tens of micrometers.

To prepare these thicker polystyrene films, the spin-coating or the Landau-Levich coating \cite{Rio2017} from a solution of polystyrene in toluene causes thickness variations due to surface instabilities triggered during the solvent evaporation \cite{Bassou2009}.
As a consequence, we prepare our polystyrene film with a solvent-free method by melting the polymer.

Glass slides are prepared with the protocol described in Sec. \ref{subsec:donor}.
Polystyrene pellets (Sigma-Aldrich, $M_w \simeq 280$ kg/mol, as used before) are placed between two glass slides, themselves placed between two aluminum plates (1 cm $\times$ 10 cm $\times$ 10 cm) held together by three screws.
This press is placed in a oven at an initial temperature of $130^\circ$C.
Screws are regularly tightened to squeeze the pellets and the temperature is raised until $220^\circ$C and the desired thickness is obtained.
The temperature is decreased to $130^\circ$C and maintained for 12 hours to relax the difference of thermal dilation between the film and the substrate and is then further decreased to the room temperature.
Aluminum plates and one of the glass slides are removed and the film is annealed again with the same protocol as described in Section \ref{subsec:psfilm}.
Films are selected for their film thickness uniformity measured with a micrometer caliper.
The film thickness is typically of $100$ $\mu$m in our experiments.

\begin{figure}
\centering
\resizebox{0.48\textwidth}{!}{
\includegraphics{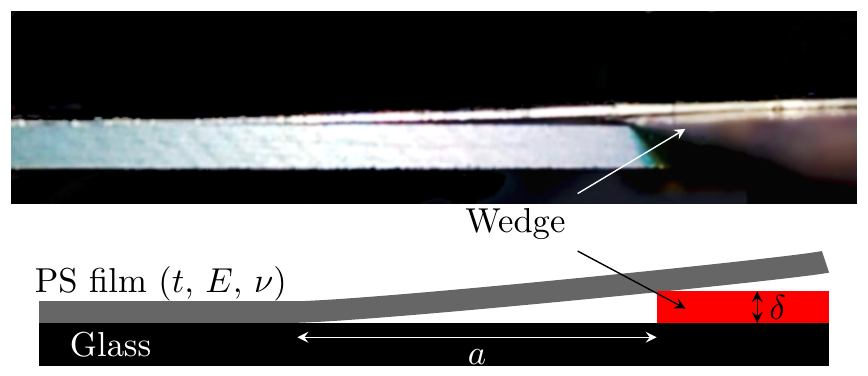}}
\caption{Picture and schematic of the cleavage test of a PS film on a glass slide.}
\label{fig:cleavage}
\end{figure}

A razor blade of thickness $\delta=385$ $\mu$m is used as a wedge.
The wedge is placed parallel and in contact with the substrate and pushed by a translational stage.
Once the crack is initiated, we stop the motor and measure the distance $a$ between the wedge and the crack tip (Fig. \ref{fig:cleavage}).

From the minimization of the sum of the bending and adhesion energies, the energy release rate is \cite{Kendall1994}
\begin{equation}\label{eq:adhesion_cleavage}
	G = \frac{3}{16} \frac{E t^3 \delta^2}{(1-\nu^2) a^4},
\end{equation}
where $\delta$ is the height of the blade and $a$ the length of the crack.
From our measurements, we obtain $G= 14\pm 2$ mJ/m$^2$, which is in good agreement with the values from the blister test presented in Table~\ref{tbl:blister_adhesion} and thus validates our approach.

\section{Conclusion}

In this work, a step-by-step guide is provided for preparation of a device to perform blister test for thin delicate materials.
More specifically, we performed blister tests to determine the adhesion energy between glass substrates and polystyrene films of micron and sub-micron thicknesses.
Side-view and top-view visualizations were performed to quantify the maximum deflection and the radius of the blister, respectively.
The adhesion energy was estimated for the experimental measurements based on the closed-form expressions proposed by Sofla \textit{et al.} (2010).
The average adhesion energy per unit area $G$ measured using the present blister test was $G = 18 \pm 2$ $ \mbox{mJ}/{\mbox m}^2$, which is in good agreement with the results of independent measurements of $G = 14 \pm 2$ $ \mbox{mJ}/{\mbox m}^2$ obtained in our cleavage test.

The current device can be used for blister test measurements in the fixed radius mode to estimate the material properties of ultra thin films.
Moreover, the protocols described here can be tailored conveniently to be applied to less conventional material pairs such as multi-layers of soft films and biomaterials such as biofilms for which performing a peeling test may damage the substrate and/or the adhering substance.

 \section{Acknowledgments}
 We thank F. Restagno and T. Salez for valuable discussions on the preparation of PS films.
F.B. acknowledges that part of the research leading to these results received funding from the People Programme (Marie Curie Actions) of the European Union's Seventh Framework Programme (FP7/2007-2013) under REA grant agreement 623541.
S.K. appreciates the early mobility funding from the Swiss National Science Foundation (P2ELP2-158896).



%
\bibliographystyle{unsrt}
\bibliography{biblio}
%
%
%

\section{Appendix}

Sofla \textit{et al.} (2010) provided the relationships for the characteristics of the film deformation \cite{Sofla2010}.
Here, we recall the expressions of the moment and the normal force.
The dimensionless moment $\bar M = \frac{(1-\nu^2) r^2 M}{E t^4}$ and the normal force $\bar N = \frac{(1-\nu^2) r^2 N}{E t^4}$ are given by
\begin{eqnarray}
\bar M &=& \frac{2}{3} \bar h + \left( \frac{m(\nu) \bar h^{1.25}}{2.2 + \bar h^{1.25}} \right) \bar h^2, \\
 \bar N &=& n(\nu) \left( -0.255 \bar h^2 \exp\left(-0.16 \bar h^{1.3}\right) + 0.667 \bar h^2\right),
\end{eqnarray}
where the dimensionless deflection is $\bar h = h/t$ and

\begin{eqnarray}
m(\nu) &=& 0.509+0.221\nu-0.263\nu^2,\\
n(\nu) &=&  0.809-1.073\nu -0.816\nu^2.
\end{eqnarray}

\end{document}